\documentclass[prb,aps,twocolumn,showpacs,superscriptaddress,floatfix]{revtex4}
\usepackage[cp1251]{inputenc}
\usepackage[english]{babel}
\usepackage{epsfig}
\usepackage{epstopdf}
\usepackage{amsmath}
\usepackage{amssymb}
\usepackage{amsfonts}
\usepackage{mathptmx}
\usepackage{eucal}
\usepackage{bm}
\usepackage{color}

\graphicspath{{Pictures1/}}

\begin{document}

\title{Density of states and current-voltage characteristics in SIsFS junctions}
\author{S.~V.~Bakurskiy}
\affiliation{Skobeltsyn Institute of Nuclear Physics, Lomonosov Moscow State University
1(2), Leninskie gory, Moscow 119234, Russian Federation}
\affiliation{Moscow Institute of Physics and Technology, Dolgoprudny, Moscow Region,
141700, Russian Federation}
\affiliation{All-Russian Research Institute of Automatics n.a. N.L. Dukhov (VNIIA),
127055, Moscow, Russian Federation}
\affiliation{MIREA - Russian Technological University, 119454 Moscow, Russian Federation}
\author{A.~A.~Neilo}
\affiliation{Faculty of Physics, Lomonosov Moscow State University, 119992 Leninskie
Gory, Moscow, Russian Federation}
\author{N.~V.~Klenov}
\affiliation{Skobeltsyn Institute of Nuclear Physics, Lomonosov Moscow State University
1(2), Leninskie gory, Moscow 119234, Russian Federation}
\affiliation{All-Russian Research Institute of Automatics n.a. N.L. Dukhov (VNIIA),
127055, Moscow, Russian Federation}
\affiliation{MIREA - Russian Technological University, 119454 Moscow, Russian Federation}
\affiliation{Faculty of Physics, Lomonosov Moscow State University, 119992 Leninskie
Gory, Moscow, Russian Federation}
\author{I.~I.~Soloviev}
\affiliation{Skobeltsyn Institute of Nuclear Physics, Lomonosov Moscow State University
1(2), Leninskie gory, Moscow 119234, Russian Federation}
\affiliation{All-Russian Research Institute of Automatics n.a. N.L. Dukhov (VNIIA),
127055, Moscow, Russian Federation}
\affiliation{MIREA - Russian Technological University, 119454 Moscow, Russian Federation}
\affiliation{Lobachevsky State University of Nizhny Novgorod, Nizhny Novgorod 603950,Russian Federation}
\author{A.~A.~Golubov}
\affiliation{Moscow Institute of Physics and Technology, Dolgoprudny, Moscow Region,
141700, Russian Federation}
\affiliation{Faculty of Science and Technology and MESA+ Institute for Nanotechnology,
University of Twente, 7500 AE Enschede, The Netherlands}
\author{M.~Yu.~Kupriyanov}
\affiliation{Skobeltsyn Institute of Nuclear Physics, Lomonosov Moscow State University
1(2), Leninskie gory, Moscow 119234, Russian Federation}
\affiliation{Moscow Institute of Physics and Technology, Dolgoprudny, Moscow Region,
141700, Russian Federation}

\date{\today }

\begin{abstract}
We study the density of states (DOS) inside superconducting Josephson SIsFS junctions with complex interlayer consisting of a thin superconducting spacer 's'  between insulator I and a ferromagnetic metal F. The consideration is focused on the local  density of states in the vicinity of a tunnel barrier, and it permits to estimate the current-voltage characteristics in the resistive state of such junctions. We study the influence of the proximity effect and Zeeman splitting on the properties of the system, and we find significant sub-gap regions with non-vanishing DOS. We also find manifestations of the 0-$\pi$ transition in the behavior of DOS in a thin s-layer. These properties lead to appearance of new characteristic features on I-V curves which provide additional information about electronic  states inside the junction.  
\end{abstract}

\maketitle

\section{Introduction}

%The superconductor-ferromagnetic hybrid structures are
%very promising for many electronic devices \cite{Eschrig1, Linder1, Blamire1,
%Soloviev1}. 

Superconductor-ferromagnetic hybrid structures are
very promising as working elements in various electronic devices \cite{Eschrig1, Linder1, Blamire1, Soloviev1}.
Starting from the first successful
experiments with $\pi$-junctions \cite{R2001, R2006}, spin-valves \cite{Baek, Gingrich},  tunable kinetic inductances \cite{BakKinInd}, and Josephson $%
\varphi$-junctions\cite{Gold1, Gold2}, 
%this topic is 
investigations of the hybrid structures became very important for fundamental studies and
possible applications \cite{RevG1, RevB1, RevV1}. Magnetic (SFS) and tunnel (SIS) junctions on the same chip \cite{Tolpygo} and even in the same stack \cite{Vernik, Nevirkovets1, Nevirkovets2, Nevirkovets3} are attractive for applications due to combination of "memory effects" and high performance.
%Moreover, the properties of electronic collectives inside such hybrid structures with thin layers can have a number of interesting features corresponding to coupling between tunnel and magnetic weak links \cite{Bakurskiy2013, Bakurskiy2016, MironovSIsFS, Bakurskiy2019}.
Moreover, the coupling via thin s-layers inside hybrid structures between tunnel barriers and magnetic weak links may result in a number of interesting
features due to interactions between superconducting states with different pairing symmetry \cite{Bakurskiy2013, Bakurskiy2016, MironovSIsFS, Bakurskiy2019}.

%Theoretical studies of the SIsFS junctions in the vicinity of 0-$\pi$ transition have already revealed non-trivial superharmonic and hysteretic current phase relations (CPR), modulation of SIs
%critical current due to proximity effect inside SFs electrode \cite{Bakurskiy2018} or
%formation of metastable energy levels \cite{Bakurskiy2017}. At the same time the
%experimental studies of SIsFS junction are mostly operating with dynamical
%measurements, i.e. current-voltage characteristic (CVC)\cite{Larkin, BakurskiyAPL, Ruppelt, Caruso1, Caruso2}.
%At the same time, the direct measurement of CPR is complicated technique and mostly considered
%for basic SFS junctions \cite{Frolov1, Frolov2}. The indirect methods of
%measuring of non-trivial CPR \cite{Baek2018} also have low application potential at that
%case, since in the wide region dynamical properties of SIsFS junction have
%trivial shape of tunnel SIS junction, and new features appears in the
%extremely narrow interval near 0-$\pi$ transition. Thus, one has lack of
%the information about the state of the middle s-layer. 

Theoretical studies of hybrid SIsFS structures in the vicinity of the 0-$\pi$  junction of its sFS part have already revealed a number of peculiarities such as nontrivial polyharmonicity and hysteresis of their current phase relationship (CPR), modulation of the SIs critical  current due to the proximity effect inside the SF electrode \cite{Bakurskiy2018}, as well as the formation of metastable energy levels \cite{Bakurskiy2017}.

Unfortunately, direct measurement of the above features is a technically difficult task \cite{RevG1, SQ}. Thus, to date, experimental determination of the CPR shape has been carried out only for solitary SFS junctions \cite{Frolov1, Frolov2, Baek2018}. As a rule, in hybrid SIsFS structures, the weak place is localized in their SIs part, so that their CPR has in most cases a trivial sinusoidal shape. 
For this reason, indirect methods of measuring non-trivial CPR \cite{Baek2018} turned out to be not very informative.
 Deviations from the sinusoidal-like CPR shape take place in a narrow range of parameters, in which the 0-$\pi$ phase transition of the junction takes place. Thus, the experimental extraction of information on the phase of the order parameter of thin s-layers is a very important problem, which allows one to draw indirect conclusions about the presence of deviations of the CPR shape from the trivial one.

\begin{figure}[t]
\center{\includegraphics[width=0.99\linewidth]{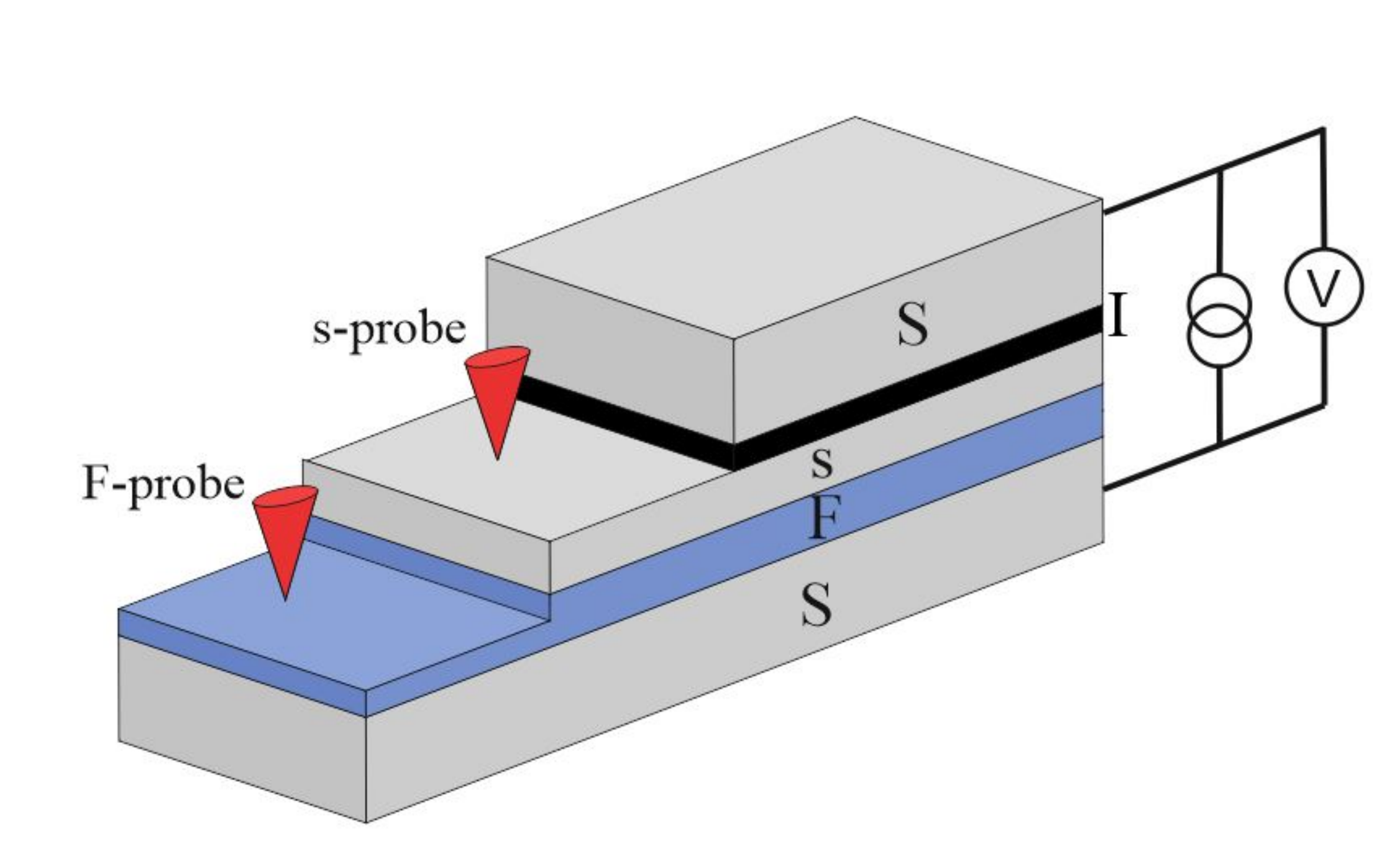}}
\caption{Sketch of the Josephson SIsFS structure with possible approach to study the DOS in F and s-films and CVC of the junction.}
\label{Sketch}
\end{figure}

In this paper we develop the method to reveal the actual state of the thin s-layer. For this purpose, we calculate the distribution of the density of states (DOS) over the structure for various parameter sets. 
%The knowledge of DOS permits to restore the behavior of the resistive part of the current-voltage characteristic (CVC)  of  hybrid SIsFS structure. 
Knowledge of DOS makes it possible to determine the behavior of the resistive part of the current-voltage characteristic (CVC) of the hybrid SIsFS structure and to correlate its features with the phase state of the s-layer.

Earlier, a similar approach was used to study the properties of hybrid structures that do not contain thin s-layers in the weak link  region \cite{BuzdinDOS,  GKF, Linder, GKS, Golikova, Tanaka, Char1, Char2, Vasenko1, Vasenko2}.
%Such approach was implemented earlier only for
%hybrid tunnel junctions without intermediate s-layer inside weak link
%\cite{BuzdinDOS,  GKF, Linder, GKS, Golikova, Tanaka, Char1, Char2, Vasenko1, Vasenko2}. 
We will demonstrate below that appearance of the addition
source of superconductivity in s-layer can significantly modify the properties of the hybrid 
junction. It should be also noted that the practical application of our approach to the study of the features of hybrid SIsFS structures is based on the study of their current-voltage characteristics, the measurement of which is a routine procedure  \cite{Larkin, BakurskiyAPL, Ruppelt, Caruso1, Caruso2}.

\begin{figure*}[t]
\center{\includegraphics[width=0.99\linewidth]{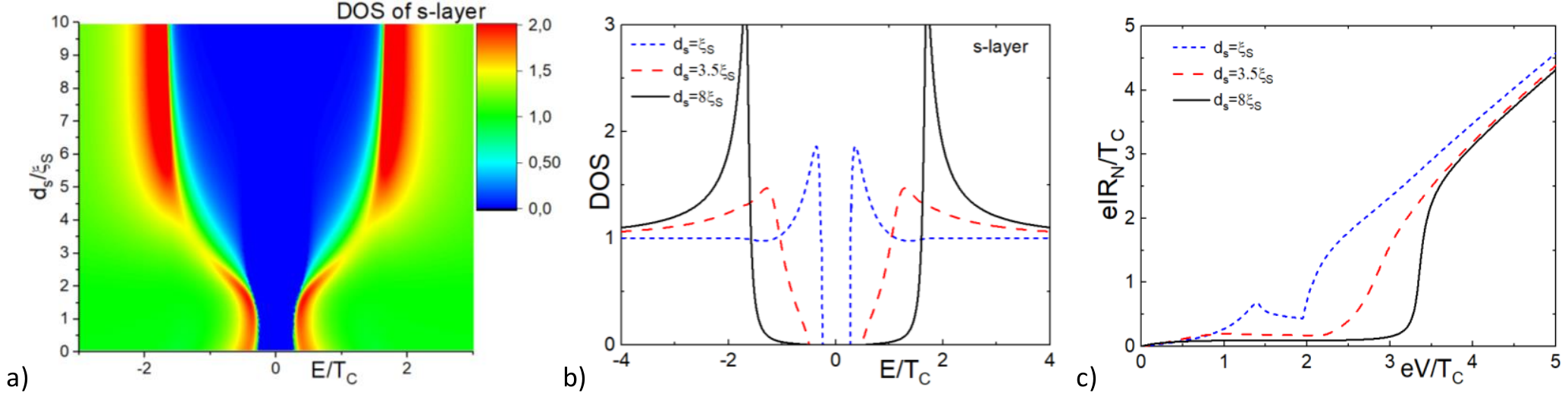}}
\caption{Local density of states at the free surface of the s-layer in sNS
structure a) as function of thickness of s-layer $d_s$ and b) for thin $%
d_s=0.1\protect\xi$ (short-dashed blue), intermediate $d_s=3.5\protect\xi$ (dashed red) and thick $d_s=9%
\protect\xi$ s-electrodes. On the panel c) the CVC of SIsNS junction are shown for the same
set of thicknesses. The other parameters are: $d_S=1\protect\xi$, $d_F=2\protect%
\xi$, $\protect\gamma_B=0.3$ and $T=0.5 T_C$. }
\label{SNS_2}
\end{figure*}

\section{Model}

The sketch of the system under consideration is shown in Fig. \ref{Sketch}: it is a conventional SIsFS junction with possibilities to study DOS in the middle of the F- layer and at the surface of the s-film. We assume, that the insulating I-layer has very low transparency, which excludes proximity effects from the top S-electrode. With this approximation, we consider the problem in the frame of Usadel equations \cite{Usadel} with Kupriyanov-Lukichev boundary conditions \cite{KL} at the inner boundaries of the 
 sFS junction  and free boundary conditions $%
\frac{d}{dx}\Phi =0$ at the free interfaces of the structure: 

\begin{equation}
\frac{\pi T_{C}\xi _{p}^{2}}{\widetilde{\omega }_{p}G_{m}}\frac{d}{dx}\left(
G_{p}^{2}\frac{d\Phi _{p}}{dx}\right) -\Phi _{p}=-\Delta _{p}  \label{fiS1}
\end{equation}%
\begin{equation}
\Delta _{p}\ln \frac{T}{T_{C}}+\frac{T}{T_{C}}\sum_{\omega =-\infty
}^{\infty }\left( \frac{\Delta _{p}}{\left\vert \omega \right\vert }-\frac{%
\Phi _{p}G_{p}}{\omega }\right) =0,  \label{delta}
\end{equation}%
\begin{equation}
\pm \gamma _{Bpq}\xi _{p}G_{p}\frac{d}{dx}\Phi _{p}=G_{q}\left( \frac{%
\widetilde{\omega }_{p}}{\widetilde{\omega }_{q}}\Phi _{q}-\Phi _{p}\right) .
\label{BC}
\end{equation}%
Here $p$ and $q$ are subscripts of corresponding layers, $G_{p}=\widetilde{%
\omega }_{p}/\sqrt{\widetilde{\omega }_{p}^{2}+\Phi _{p,\omega }\Phi
_{p,-\omega }^{\ast }},$ $\widetilde{\omega }_{p}=\omega +iH_{p}$, $\omega
=\pi T(2n+1)$ are Matsubara frequencies, $\Delta _{p}$ is pair potential , $%
H_{p},$ is exchange energy of the ferromagnetic layer ($H_{p}=0$ in
nonferromagnetic materials), $T_{C}$ is critical temperature of
superconductor, $\xi _{p}=(D_{p}/2\pi T_{C})^{1/2}$ is coherence length, $%
D_{p}$ is diffusion coefficient, $G_{p},$ and, $\Phi _{p},$ are normal and
anomalous Green's functions, respectively, $\gamma _{Bpq}=R_{Bpq}\mathcal{A}%
_{Bpq}/\rho _{p}\xi _{p}$, is suppression parameter, $R_{Bpq}$ and $\mathcal{%
A}_{Bpq}$ are resistance and area of corresponding interface. The sign plus
in (\ref{BC}) means that $p$-th material is located at the side $x_{i}-0$
from interface position $x_{i}$, and sign minus corresponds to the case where
$p$-th material is at $x_{i}+0$. 

In order to determine DOS, at the first stage, we have numerically solved
the set of equations (\ref{fiS1})-(\ref{BC})
and have determined the spatial distribution
 of the self-consistent pair potential  $\Delta (x)$
 across the whole structure. We have simplified the problem assuming 
 that the coherence lengths, $\xi_S = \xi_F = \xi$, 
 and the resistivities, $\rho_S = \rho_F = \rho$, of superconducting and ferromagnetic materials are equal to each other.  
At the second stage, in order to obtain the required  
%Then we solve
%the problem for determined $\Delta (x)$ in the real energy approach.  It
%consists of the similar set of equations (\ref{fiS1}) - (\ref{BC}) with change
%of Matsubara frequency to real energies: $\omega =-\iota E$ and provides
spatial distribution of DOS $N(E,x)=Re G(E,x)$,
we have analytically continued the equations (\ref{fiS1}) - (\ref{BC})  by passing from the Matsubara frequencies to the real energy $\omega =-\iota E$
and numerically solved the resulting system of equations using the function $\Delta (x)$ determined at the first stage.

\begin{figure*}[t]
\center{\includegraphics[width=0.99\linewidth]{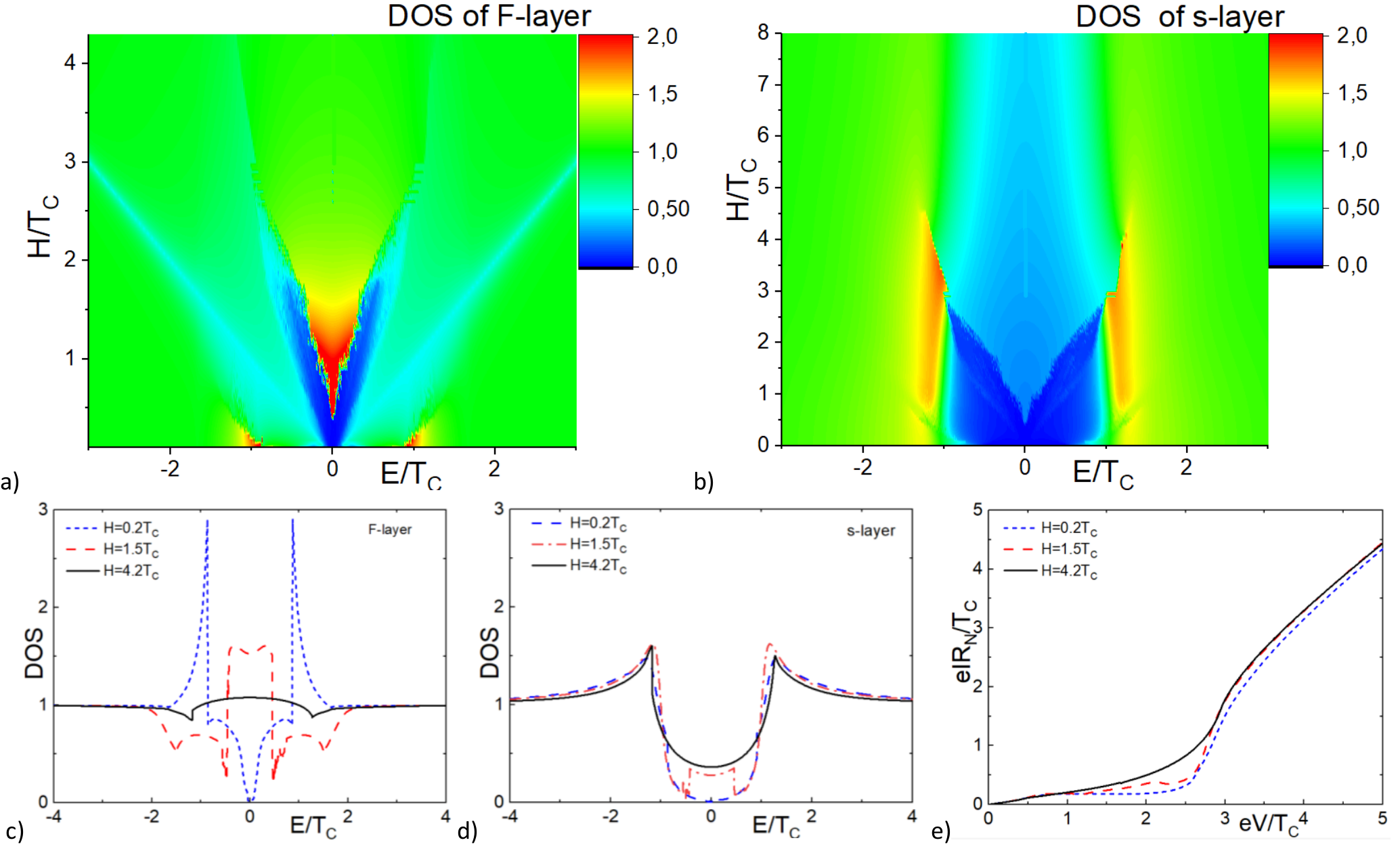}}
\caption{ DOS of sFS structure with intermediate thickness of F and s-layer ($d_F=2.0\xi$, $d_s=3.5\xi$) a) in the middle of F-layer and b) at the surface of the s-layer as function of exchange energy H; c,d) DOS in the middle of the F-layer c) and at the free surface of the s-layer d) for small $H=0.2T_C$ (blue short-dashed line), intermediate $H=1.5T_C$ (red dashed line), and large $H=4.2T_C$ (black solid line) exchange energies; e) IV-curves of the related SIsFS junction with small, intermediate and large exchange energies.  The other parameters are $\protect\gamma_B=0.3$ and $T=0.5 T_C$.  }
\label{Hintint}
\end{figure*}

In the limit of small I- layer transparency the knowledge of the local DOS at the free surface of the thin s-layer provides a possibility to calculate the current-voltage characteristic of the
tunnel SIsFS junction using the tunnel formula \cite{Werthamer}%
\begin{equation}
I=\frac{1}{eR_{N}}\int\limits_{-\infty }^{\infty
}N_{1}(E-eV)N_{2}(E)[f(E-eV)-f(E)]dE,
\end{equation}
%in assumption of small transparency, which excludes the
%proximity effect induced from the left electrode. 
Here $V$ is a voltage drop across the tunnel layer,  $%
R_{N}$ is normal resistance of the tunnel barrier,  and  $f(E)=(1+exp
(E/T))^{-1}$ is equilibrium Fermi distribution function.

\section{Density of States in SI\lowercase {s}NS structure}

%Initially we consider the density of state of the similar
%SNs junction without exchange in the metallic layer of the weak link and create a reference point. 

To get the reference point we have started with investigation of SIsNS structure in which the exchange energy is zero in the non superconducting layer. We calculate local density of states on the free surface of s-layer as a function of its thickness $d_s$ (see Fig. \ref{SNS_2}a,b) which has the shape of a cup. At $d_s \gtrsim 5\xi$ 
 the proximity effect with the normal layer weakly affects the density of states. In this interval of thicknesses, both the order parameter and the anomalous Green's functions reach their bulk values near the Is interface. As a result, DOS
 has the BCS form with peaks at the energies equal to the bulk value of the order parameter, $\Delta_0,$ and the minigap, $\Delta_1,$ at $E \le \Delta_0.$ Typical  behavior of DOS calculated for $d_s=8\protect\xi$ is presented in  Fig. \ref{SNS_2}b by the black curve.  For $d_s<5\xi$ 
the proximity effect leads to noticeable deviations of the anomalous Green's functions from bulk values, providing their spatial dependence on x coordinate. This is accompanied by the decrease of the minigap magnitude and the smearing of the singularity in the density of states (shown by the dashed red curve in Fig. \ref{SNS_2},b calculated for $d_s=3.5\protect\xi$ ). 
A further decrease in $d_s$ thickness is accompanied by the restoration of the singularity in the density of states at an energy equal to the minigap (see the dashed short-dashed blue in Fig. \ref{SNS_2}b calculated for $d_s=0.1\protect\xi$ ). Previously, such transformation in the density of states with decreasing s-layer thickness was predicted in the sN bilayer  \cite{Sov}. Physically the effect follows from the fact that the smaller $d_s$ the larger the energy interval in a vicinity of the minigap in which both an order parameter and anomalous Green's functions restore their independence on spatial coordinate $x$. The situation turns out to be similar to that realized at large thicknesses $d_s>5\xi$ up to replacement of $\Delta_0$ by the minigap. Note that at $d_s=0.1\protect\xi$ 
 there is no intrinsic superconductivity in the sN bilayer. However, in the sNS part of the SIsNS structure, superconducting correlations penetrate into the s-layer from the massive S electrode, thereby maintaining superconductivity of this layer.
This leads to the shape of the density of states significantly different from the one calculated for the SINS structure \cite{SINS}
 with the same thickness of the N-layer as that of the sN bilayer, but with the effective constant of the electron-phonon interaction equals to zero.

\begin{figure*}[t]
\center{\includegraphics[width=0.99\linewidth]{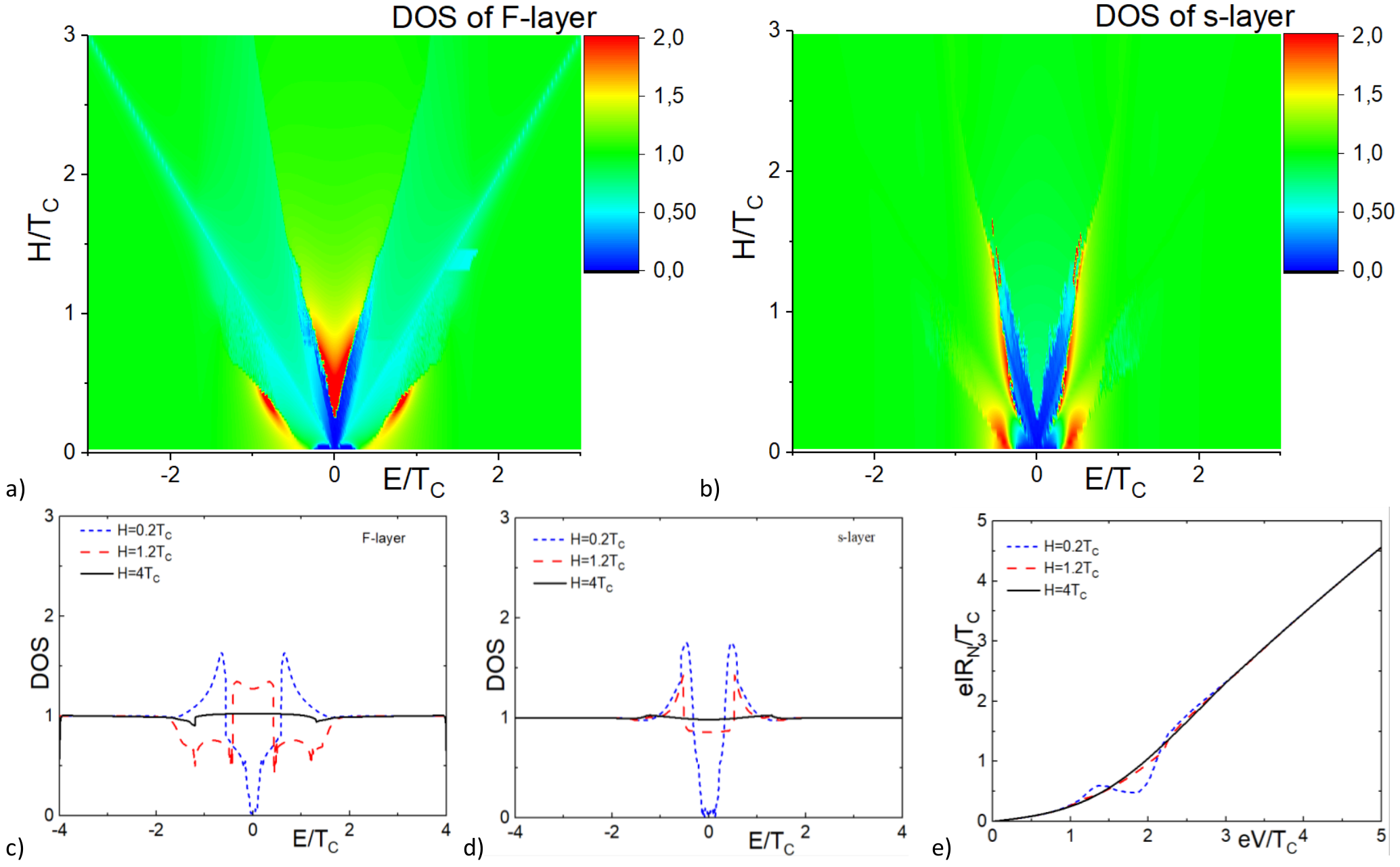}}
\caption{ DOS of sFS structure with thin s-layer ($d_s=\xi$) and intermediate thickness of F-layer  $d_F=2.0\xi$  a) in the middle of F-layer and b) at surface of s-layer as function of exchange energy $H$; c,d) DOS in the middle of F-layer c) and at the free surface of s-layer d) for small $H=0.2T_C$ (blue short-dashed line), intermediate $H=1.2T_C$ (red dashed line), and large $H=4.2T_C$ (black solid line) exchange energies; e) IV-curves of the related SIsFS junction with small, intermediate, and large exchange energies.   The other parameters are $\protect\gamma_B=0.3$ and $T=0.5 T_C$.   }
\label{Hintsm}
\end{figure*}

Figure \ref{SNS_2}c shows the manifestation of the above mentioned DOS features in the I - V characteristics of the junctions. As expected, in the case of thick
s-layer the standard tunnel I-V characteristics are realized. In the intermediate thickness range ($d_s=3.5\xi$) the broadening of the DOS leads to the decrease of the slope of CVC. In the limit of small $d_s$, the additional peak feature at $V=\Delta_0-\Delta_1$ appears, as it should be for the SIS' tunnel devices with different superconducting electrodes. 
\section{Density of States in SI\lowercase {s}FS structure}

The existence of ferromagnetic ordering in the non-superconducting region of the SIsFS structure enhances the effect of suppressing superconductivity in the thin s-layer. Below we will restrict ourselves to consideration of two cases. 
We begin with an analysis of what changes in the shapes of the DOS and I -V characteristics take place with an increase in the exchange energy in the F- region of the SIsFS structure for thin $d_s=\xi$ (Sec.\ref{ThickF1})  and intermediate $d_s=3.5\xi$ (Sec.\ref{ThickF2}) superconducting s-layer.
Then, in  the Secs.\ref{ThinF1} and \ref{ThinF2}, we will study what happens to the occuring features at thinner ferromagnetic thicknesses  $d_F=0.5\xi$.

%The next considered feature in the Sec.\ref{ThickF} is the influence of the exchange field on the system. We study it with several sets of parameters: initially we take the same thickness of F layer as in the latter case $d_F=2\xi$ and study the dependence of DOS versus exchange energy $H$ for thin $d_s=\xi$ and intermediate $d_s=3.5\xi$ superconducting s-layer. However, in the case of the large exchange energies the proximity effect was suppressed too strong, so we also study the properties for thinner F-layer  $d_F=0.5\xi$ in the Sec.\ref{ThinF}. 

\subsection{Thick F-  and intermediate s- layers, $d_F=2\xi$,  $d_s=3.5\xi$ \label{ThickF1}}

Figures \ref{Hintint}a,b show DOS of sFS structure  as function of exchange energy $H$ calculated for intermediate thickness of the F and s-layer ($d_F=2.0\xi$, $d_s=3.5\xi$) in the middle of the F-layer (Fig. \ref{Hintint}a) and at the surface of the s-layer (Fig. \ref{Hintint}b). 

Fig. \ref{Hintint}a shows that an increase in $H$ is accompanied by a Zeeman splitting of the peak in the DOS located at  minigap and a shift of one of them in the direction to the Fermi energy ($E=0$) resulting in decreasing the mini gap magnitude. At $H=0.2T_C$ (see blue short-dashed line in Fig. \ref{Hintint}c) the mini gap still exists and splitting of the peak is not yet resolved.  At $H=1.5T_C$ (see  red dashed line in Fig. \ref{Hintint}c)  the mini gap disappears and one of the splitted peaks crosses the  Fermy energy resulting in formation of a DOS plateau in its vicinity. Finally,  at larger exchange energies $H>2T_C$ (see  black solid line in Fig. \ref{Hintint}c) the energy interval between the splitted peaks significantly exceeds $\Delta_0$, and the singularities in the DOS practically disappear in the subgap energy range.

The structure of the DOS at the free surface of the s-layer contains significantly fewer features than the DOS in the middle of the F layer. At the same value of $H$ the width of the energy gap in DOS on the top of  s-layer (see the area colored blue in Fig.\ref{Hintint}a and Fig.\ref{Hintint}b is around $T_C$, which is significantly larger than the gap inside the DOS in the F-layer. 
At $H \approx 0.5$ the gap in DOS is closed resulting in nucleation of a plateau. For the same $H$  it is wider than in the F- film, but its amplitude is several times smaller than that in the F-layer. 
At $H \approx 3T_C$ the gap in the s-layer DOS  is closed, although even at the larger $H$ there is a significant wide decrease of the DOS in a vicinity of zero energy. 

\begin{figure*}[t]
\center{\includegraphics[width=0.99\linewidth]{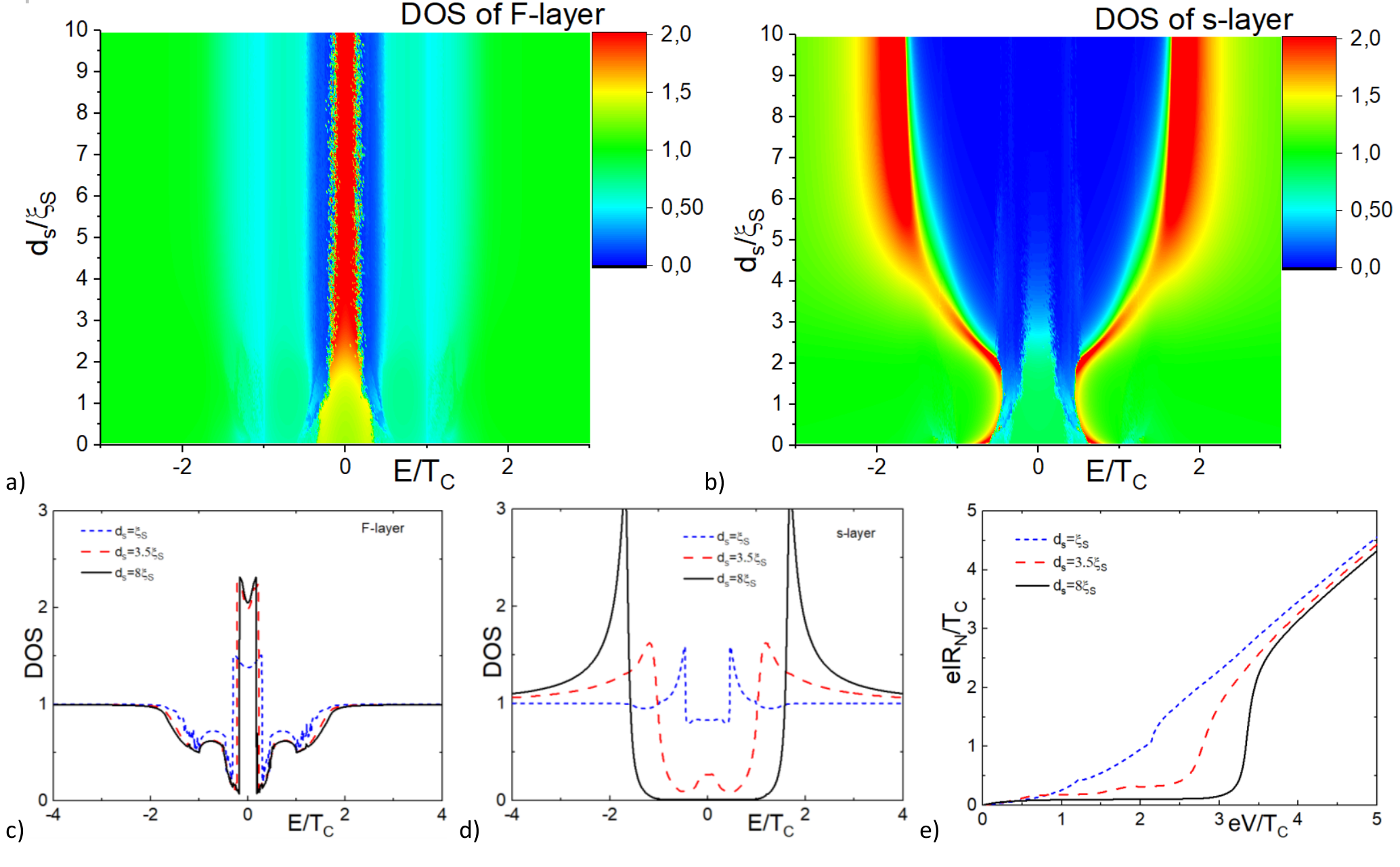}}
\caption{ DOS of sFS structure with intermediate thickness of the F-layer ($d_F=2.0\xi$) and exchange energy $H=T_C$ a) in the middle of the F-layer and b) at the surface of the s-layer versus its thickness $d_s$; c,d) DOS in the middle of the F-layer c) and at the free surface of the s-layer d) for small $d_s=1\xi$ (blue short-dashed  line), intermediate $d_s=3.5\xi$  (red dashed line), and large $d_s=8\xi$  (black solid line) exchange energies; e) IV-curves of the related SIsFS junction with small, intermediate and large thickness of the s-layer.   The other parameters are $\protect\gamma_B=0.3$ and $T=0.5 T_C$.   }
\label{ds_int_1}
\end{figure*}

The panels c,d in Fig. \ref{Hintint} show the cross-sections of the studied density of states in the F- (c) and s- (d) layers for small $H=0.2T_C$ , lower-intermediate $H=1.2T_C$ and upper-intermediate $H=4.2T_C$ exchange energies. 
It is seen that 
at $H=0.2T_C$ (see blue short-dashed line in Fig. \ref{Hintint}d) the shape of the DOS is close to that presented by red dashed curve  in Fig. \ref{SNS_2}b. The only distinction  is in smaller minigap value. At $H=1.5T_C$ (see  red dashed line in Fig. \ref{Hintint}d) there is a plateau in DOS in the same energy interval as that in the F-layer DOS. But the amplitude of this peculiarity is approximately three times smaller than  in  DOS inside of the F-film. 
For $H=4.2T_C$, the difference between DOS in the F- and s-layers is most significant. While in the F- film the subgap DOS has practically no features, the DOS at the surface of the s-layer still exhibits the peaks at an energy around $\Delta_0$ and there is a significant dip in the density of states at low energies.

\begin{figure*}[t]
\center{\includegraphics[width=0.99\linewidth]{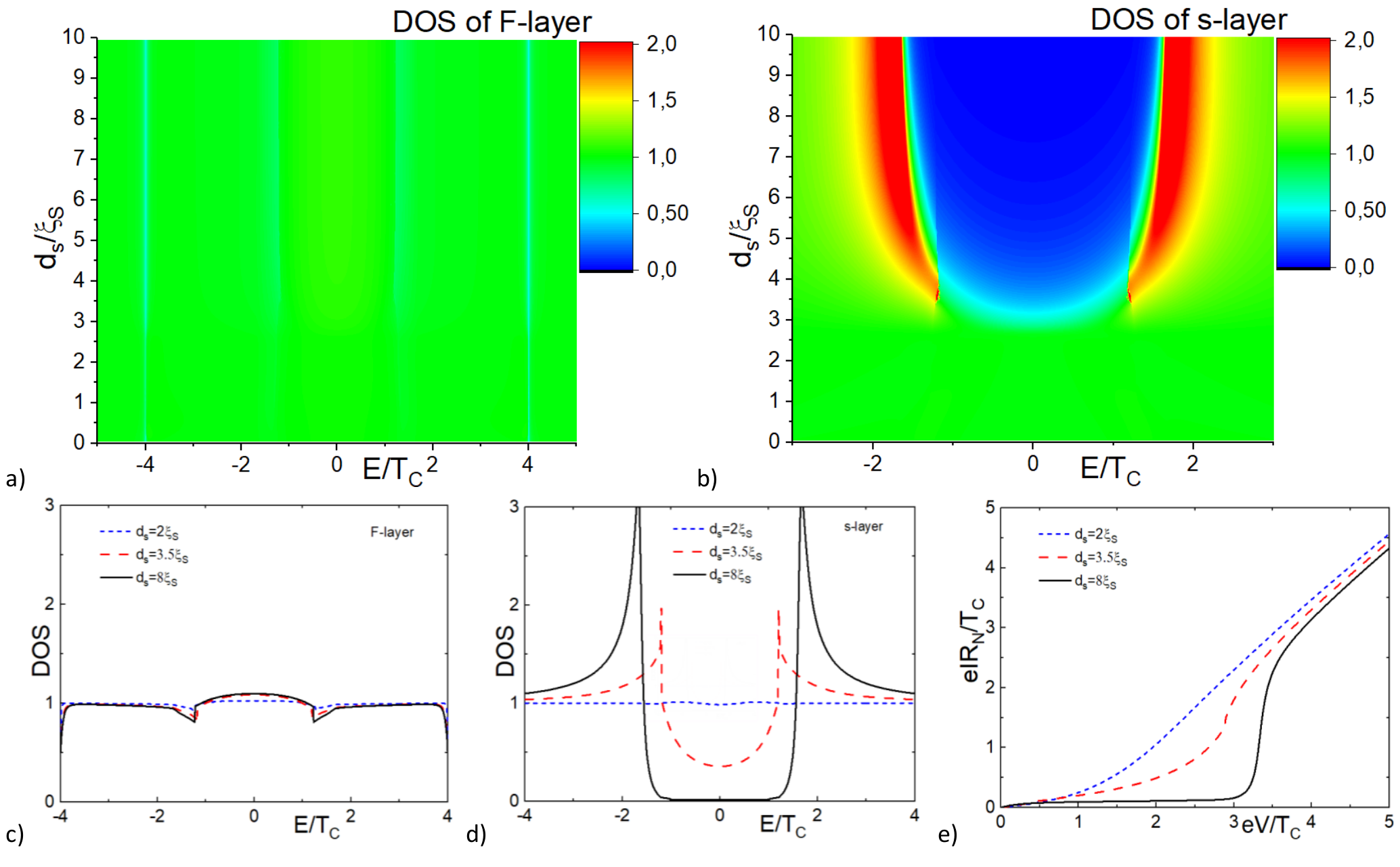}}
\caption{ DOS of sFS structure with intermediate thickness of the F-layer ($d_F=2.0\xi$) and exchange energy $H=4T_C$ a) in the middle of the F-layer and b) at the surface of the s-layer versus its thickness $d_s$; c,d) DOS at the middle of the F-layer c) and at the free surface of the s-layer d) for small $d_s=2\xi$ (blue short-dashed  line), intermediate$d_s=3.5\xi$  (red dashed line), and large $d_s=8\xi$  (black solid line) exchange energies; e) IV-curves of the related SIsFS junction with small, intermediate and large thickness of the s-layer.   The other parameters are $\protect\gamma_B=0.3$ and $T=0.5 T_C$.   }
\label{ds_int_4}
\end{figure*}

%The results of the first case with intermediate thickness of s and F-layers $d_s=3.5\xi$, $d_F=2.0\xi$ are shown in the Fig. (\ref{Hintint})
%The panel in Fig.3a shows DOS inside F-layer.

% At small exchange energies, it has a small gap inside, the increase of $H$ leads to splitting of the gap in two different areas with large plateau between them due to Zeeman effect. At the same time the coherence peaks on s-layer DOS at $E=\Delta$  almost disappear. At the larger exchange energies ($H>2T_C$) the splitted gaps disappear too. 

%The similar features can be found on the DOS of the s-layer (Fig.(\ref{Hintint}b). However, in superconductor the value of the gap is determined by the self-consistent properties of that layer, and the width of the gap is around $T_C$, which is significantly larger than the gap inside the F-layer. 

%Sub-gap zone also appears in the DOS of the s-layer with increase of exchange field at $H=0.5$, but the amplitude of DOS inside this zone is much smaller and varies in the interval $(0.3 T_C , 0.5 T_C)$. The panels c,d in Fig. \ref{Hintint} show the cross-sections of the studied density of states in F (c) and S (d) layers for small $H=0.2T_C$ , lower-intermediate $H=1.2T_C$ and upper-intermediate $H=4.2T_C$ exchange energies. 

The presence of the subgap zone with non-vanishing DOS leads to appearance of additional features at IV-curves of the corresponding SIsFS structure (See Fig.3e). 
It is seen that at  the intermediate exchange field (see the red curve in Fig.3e calculated for   $H=1.5T_C$) there is additional peak on CVC at $eV \approx  2T_C$, which corresponds to the sum of the gap $\Delta_0$ of the bulk  S electrode and plateau energies. 
This feature starts to manifest itself from the moment of the formation of a plateau on the DOS at $H \approx 0.2 T_C$  and is broadened with an increase in H. It disappears along with the disappearance of the features that form this plateau in DOS and does not exist at  $H > 4.2 T_C$.

%At the same time such peak doesn't appear in the limits of small and large exchange energies. That peak is strongly dependent from the actual value of the effective exchange field. So in the real weak ferromagnets (they are usually alloys of normal and ferromagnetic metals) it is distinguished from the similar differential peak of SIsNS junction with extremely high sensitivity to remagnitization of the F-layer, which changes the domain or cluster structure and modifies effective exchange, averaged on the coherence length $\xi$. 

\begin{figure*}[t]
\center{\includegraphics[width=0.99\linewidth]{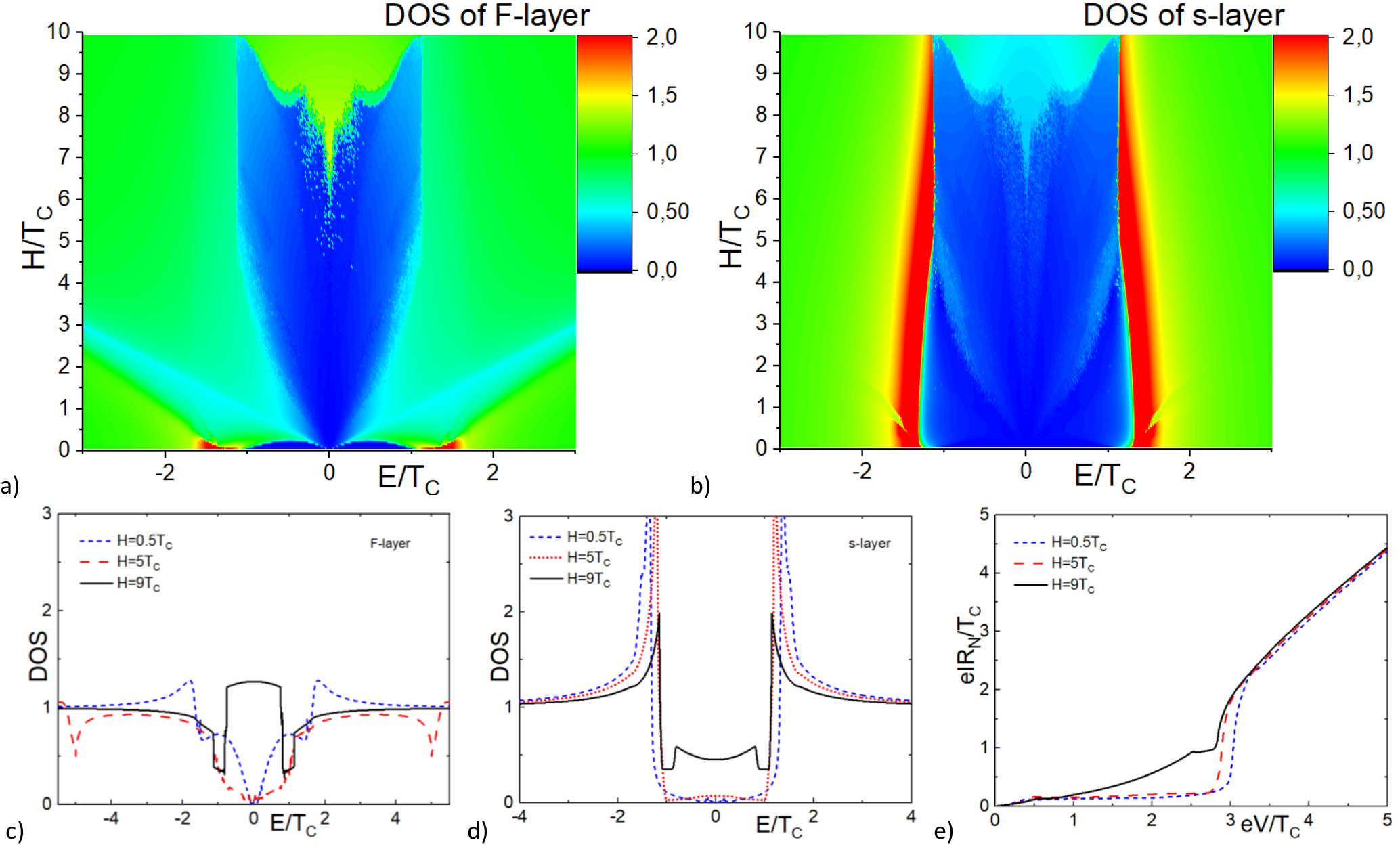}}
\caption{ DOS of sFS structure with thin F-layer and  intermediate thickness of the s-layer ($d_F=0.5\xi$, $d_s=3\xi$) a) in the middle of the F-layer and b) at the surface of the s-layer as function of exchange energy H; c) at the free surface of the s-layer d) for small $H=0.5T_C$ (blue short-dashed line), intermediate $H=5 T_C$ (red dashed line) and  large $H=9T_C$ (black solid line) exchange energies; e) IV-curves of the related SIsFS junction with small, intermediate and large exchange energies.  The other parameters are $\protect\gamma_B=0.3$ and $T=0.5 T_C$. }
\label{H_sm_int}
\end{figure*}

\subsection{Thick F- and thin s- layers, $d_F=2\xi$,  $d_s=\xi$ \label{ThickF2}}

%With such small thicknesses of the s-layer, intrinsic superconductivity is guaranteed to be absent in it. 
At $d_s=\xi$ there is no intrinsic superconductivity in the s-layer. 
Therefore the shape of the DOS inside the F-layer is completely determined by superconducting correlations induced into the F film from the bulk superconducting electrode (see Fig. \ref{Hintsm}a,c) and is rather close to that presented in Fig. \ref{Hintint}a,c.
%In the Fig. \ref{Hintsm}  we demonstrate the properties versus exchange energy $H$ in the case of the thinner s film with thickness $d_s=\xi$. The properties of the DOS inside the F-layer are almost the same.

The main peak disappears around exchange energy $H=0.5 T_C$, while the local minimum on the DOS is growing linearly with increase of $H$. The minigap becomes splitted at $H=0.25 T_C$ with appearance of the middle peak which transforms into plateau at larger exchange energies. It is interesting, that on the DOS($H$) dependence there are two different features with dips on it. The first one appears at $E=H$ due to Zeeman splitting of main peaks of DOS. The second one also has linear dependence versus H with smaller coefficient and correlate with the limits of mini-gap inside the F-layer.

Since s-layer is thin, the order parameter and Green's function inside it are nearly constant in a wide interval of energy. 
%and acts as normal metal layer,
For this reason  the DOS at the surface of the film  doesn't significantly differ from the DOS of the N-layer (see short-dashed blue curve in Fig. \ref{SNS_2}b). With $H$ increase the gap is closed, the amplitude of the peaks suppressed, while their width increases. 
%However, while the zero-energy zone in F-layer is higher than unity, on the surface of s-layer it doesn't reach it at all. We also find, that coherent peak have non-trivial behavior versus $H$ on the surface of s-layer. At small exchange energies $H<0.5 T_C$ it appears in the same position $(E\approx0.5 T_C)$ in the F- and s-layers, and probably has the origin from the coherence peak of the large S- electrode. The position of this peak is smaller than the bulk value of $\Delta$, since the pair potential of the S-electrode near SF interface is also suppressed due to inverse proximity effect. At larger exchange energy this peak disappears, however another peak appears exactly at the edge of the effective mini-gap.

At  $H> 0.5 T_C$ the obtained behavior of DOS leads to the resistive branch of CVC similar to that of the NIS junctions presented in Fig. \ref{Hintsm}e. The minor difference from NIS junction is provided by deviations of DOS in the vicinity of the minigap (small increase above and small decrease below). At rather small exchange energies, the CVC becomes hysteretic (in the case of voltage biasing) with wide maximum 
at voltage in the vicinity of difference between bulk gap and gap of the s-layer.

\begin{figure*}[t]
\center{\includegraphics[width=0.99\linewidth]{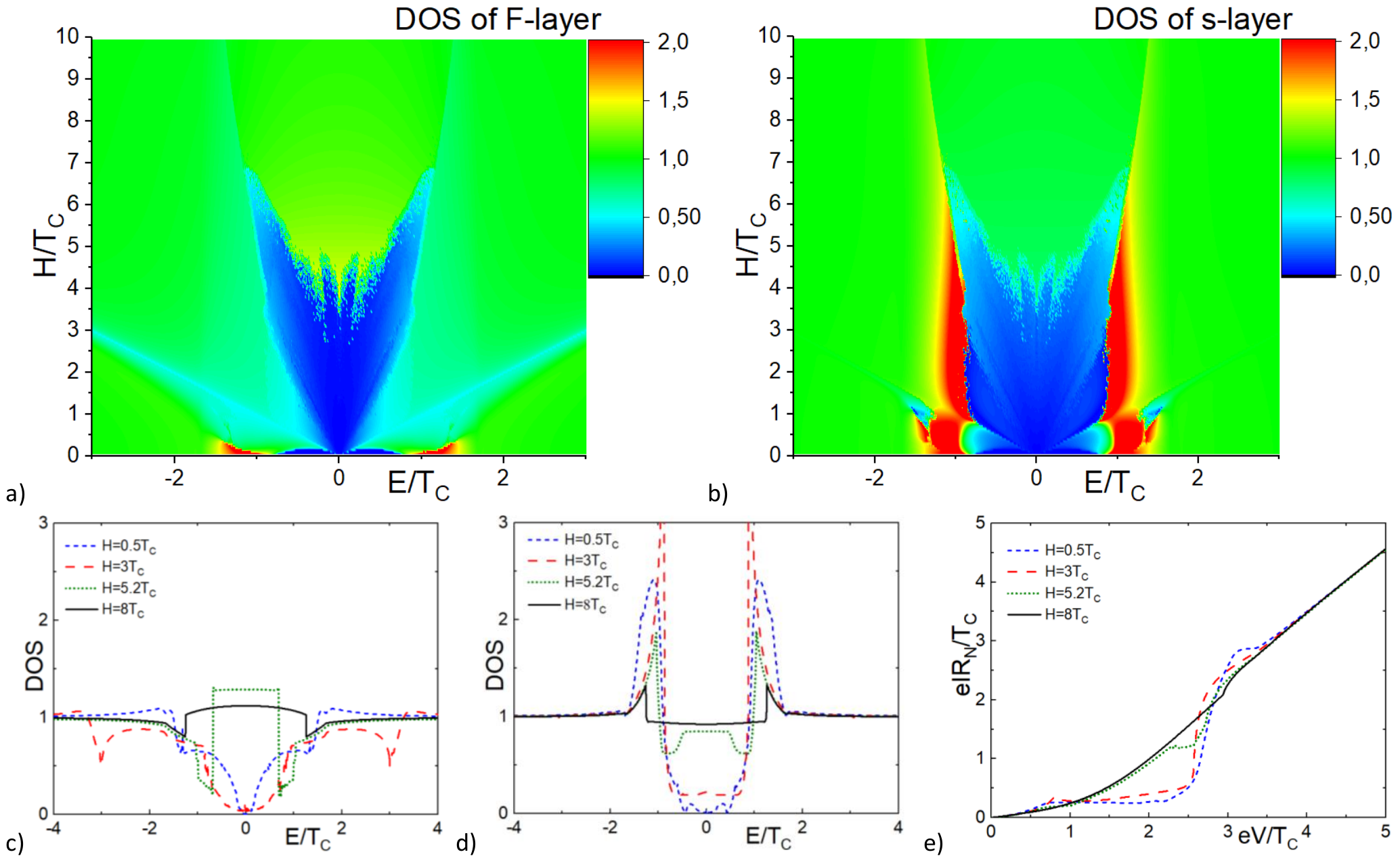}}
\caption{ DOS of sFS structure with thin thickness of the F- and s-layer ($d_F=0.5\xi$, $d_s=0.6\xi$) a) in the middle of the F-layer and b) at the surface of the s-layer as function of exchange energy H; c,d) DOS in the middle of the F-layer c) and at the free surface of the s-layer d) for small $H=0.5T_C$ (blue short-dashed line), intermediate $H=3.T_C$ (red dashed line), upper-intermediate $H=5.2T_C$ (green  dotted line), and  large $H=8T_C$ (black solid line) exchange energies; e) IV-curves of the related SIsFS junction with small, intermediate, and large exchange energies.  The other parameters are $\protect\gamma_B=0.3$ and $T=0.5 T_C$.  }
\label{H_sm_sm}
\end{figure*}

\subsection{Thick F-layer, $d_F=2\xi$.  Evolution of DOS with $d_s$ increase.  \label{ThickFH}}

Figures \ref{ds_int_1}a,b and Fig.\ref{ds_int_4}a,b show the evolution of DOS with increase of $d_s, $  calculated for  $H=T_C$ and $H=4T_C,$ respectively. 
%The dependence of the surface DOS versus $d_s$ supports the statements above at Fig.\ref{ds_int_1} and Fig.\ref{ds_int_4} at finite exchange energies $H = 1 T_C$ and $H=4 T_C$ respectively. 
The 
%general view 
shape
of the DOS versus $d_s$ on the top of the s-layer is similar to DOS of SNs junction (see. Fig. \ref{ds_int_1}b). The cup shapes are  due to the same processes of reappearance of the intrinsic superconductivity in the s-layer.  

The difference in DOS between the SFs and SNs structures is in the formation of the zero-energy peak inside the gap at the cup base.  At $H=T_C$ the peak nucleates due to violation of time-reversal symmetry in the F layer. It occurs for all $d_s$ in ferromagnetic layer, but disappears in the s-layer at $d_s>4\xi$, when the intrinsic superconductivity restores in it. At the same time, in the region of parameters, where zero-energy peak exists in the s-layer, the amplitude of the peak inside the F-layer is significantly smaller and the peak becomes wider than at large $H$ values.

Such nontrivial DOS behavior is a feature of weak ferromagnets. Even at $H=4T_C$, the density of states in the F-layer becomes independent versus thickness of the s-layer and has the properties of normal metal DOS (Fig. \ref{ds_int_4}a,c). In the superconductive layer it also has trivial properties of normal metal at the thickness smaller than the critical one ($d_s<3.5\xi$) and has a conventional gap at the larger thicknesses (Fig. \ref{ds_int_4}b,d) . 

Regarding the properties of CVC of corresponding SIsFS junction, the change of $d_s$ leads to the transition from SNS like IV-curve at large $d_s$ to SIN-like dependence for smaller thickness (Figs. \ref{ds_int_1}e and \ref{ds_int_4}e).  However, the features appearing in the case of small exchange energies exhibit some additional jumps on IV-curves of structures with thin s-layer, which correspond to the presence of narrow gaps on the sides of zero-energy DOS peak. 

\begin{figure*}[t]
\center{\includegraphics[width=0.99\linewidth]{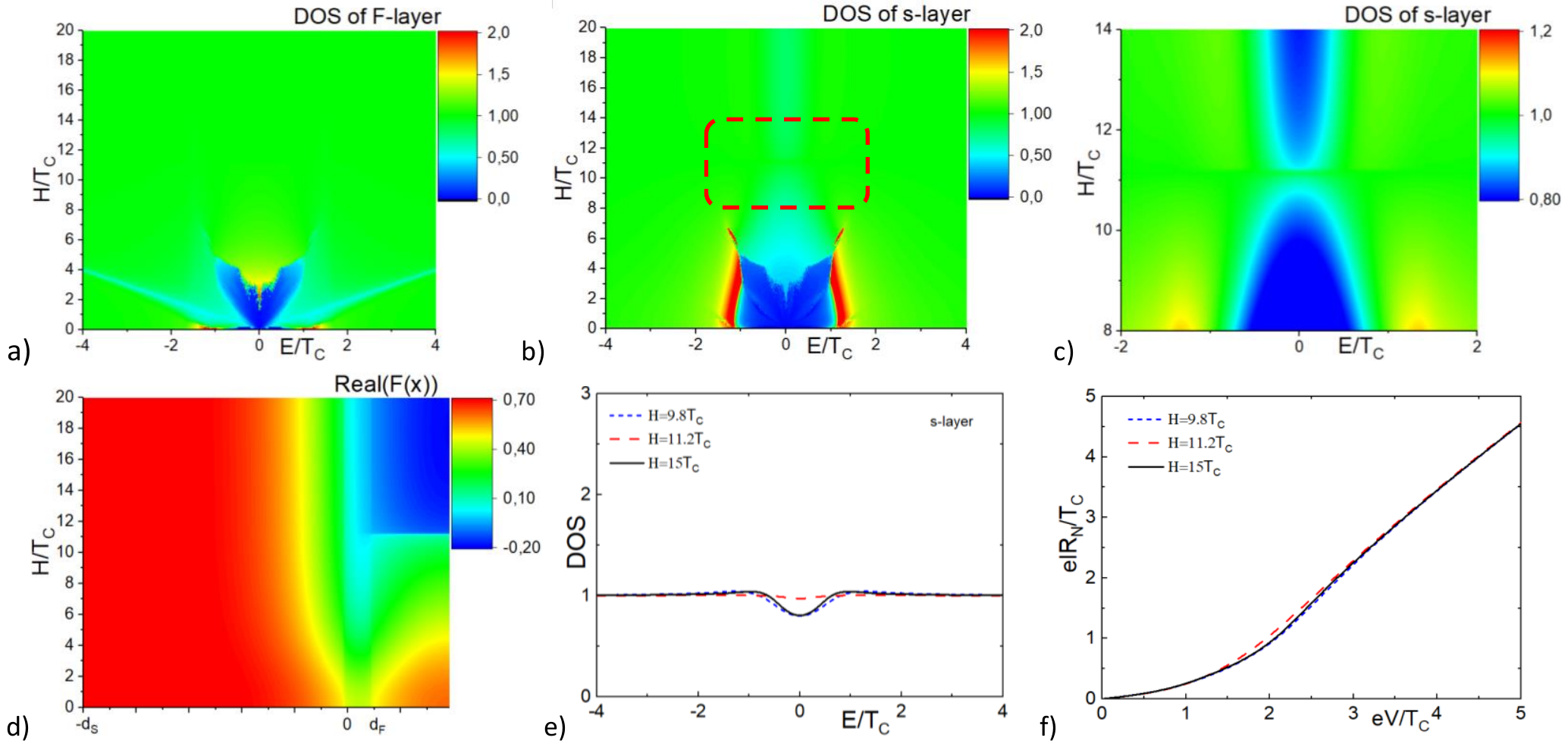}}
\caption{ The electronic properties of SIsFS junction with $d_F=1\xi$, $d_s=3\xi$, $\gamma_B=0.3$ at $T=0.5 T_C$ for different exchange energy $H$. The top panels show DOS a) in the middle of the F-layer and b) at the surface of the s-layer; c) enlarged DOS of the s-layer in the region of 0-$\pi$ transition marked by dashed line on panel b); d) the corresponding spatial distribution of pair amplitude $F_1(x,H)$ calculated on the 1st Matsubara frequency, e) the DOS at the surface of the s-layer in the 0-state (blue short-dashed line), $\pi$-state (black solid line) and inside 0-$\pi$ transition (red dashed line). f) IV-curves of the related SIsFS junction inside 0, $\pi$ and transitioning state. The other parameters are $\protect\gamma_B=0.3$ and $T=0.5 T_C$.  }
\label{0pitr}
\end{figure*}

\subsection{Thin F- and intermediate s- layers , $d_F=0.5\xi$, $d_s=3\xi$  \label{ThinF1}}

%\begin{figure}[t]
%\center{\includegraphics[width=0.99\linewidth]{L_F=0p5p1.pdf}}
%\caption{ Local density of states on the free surface of s-layer  as function of thickness of s-layer $d_s$ in sNS
%structure with thin N-layer $d_N=0.5\xi$}
%\label{DOS_S_F05}
%\end{figure}

%The above results correspond to the case of thick enough F-layer $d_F=2\xi$. At the same time, such choice of the parameters provides strong suppression of proximity effect in the case of large $H$. 
To study possible manifestation of penetrating proximity effect between bulk S and thin s-layers in this section we study 
%similar properties 
DOS and CVC of the SIsFS  system at $d_F=0.5\xi$. 

As above,  we have begun with study the DOS and CVC of the referent SIsNS junction by putting  $H=0$ in the F film. The calculated DOS  has  the form similar to that presented in Fig.1 with a wider base of the cup-shaped structure.
%However, the dependencies versus exchange energy are drastically different from the previous case. 
However, the transformations of the DOS and CVC shapes with an increase in the exchange energy are fundamentally different from the cases considered above.
Figure \ref{H_sm_int} demonstrates the characteristics of the SIsFS structure with intermediate thickness of the s-layer $d_s=3\xi$. From Fig. \ref{H_sm_int}a it follows that at rather small $H$  there is 
%In this case, the low-exchange state ($H<0.5T_C$) is more significant on the DOS of F-layer.
the wide gap in DOS in the F film. It nucleates due to proximity effect with the S-layers. The gap is rapidly closing with increase of exchange field. At $H=5 T_C$ it tends to reopen to a width  $\approx T_C$. Simultaneously  at the same  exchange energy $H=5 T_C$ the zero-energy feature also nucleates and broadens with further increase of the $H$. At $H=10T_C$ zero-energy zone becomes so wide, that it totally closes the gap again. 
%Unfortunately in this region of parameters algorithm of DOS calculation becomes very unstable, providing some broken points, which are rippling the middle part of the figure. 

Inside the s-layer the gap in DOS is nearly constant (see Fig.\ref{H_sm_int}b) at $H \lesssim 10T_C$. Only its  
 little decrease is observed with the growth of exchange energy. It happens since thin ferromagnetic layer doesn't effect strongly on the formation of pair potential inside the s-layer at $H \lesssim 10T_C$. 

Also stronger proximity effect  is leading to appearance of the feature on the main peak of the DOS of the s-layer at $H\approx T_C$. At that exchange energy, the Zeeman dip feature  in the F-layer DOS intersects position of the coherence peak in the superconductor. Zeeman dip feature is induced to the s-layer DOS due to proximity, providing the local splitting of the main coherence peak in the vicinity  $H\approx T_C$  in  Fig.\ref{H_sm_int}b .

Also at high exchange energies,  the subgap zone also forms in the DOS of the s-layer  $H>7 T_C$   in  Fig.\ref{H_sm_int}b . The contour of that zone is almost the same with F-layer's one, but its amplitude is significantly smaller. Thus, in the case of thin F-layer and intermediate s-layer, the superconducting layer is protected enough to demonstrate its intrinsic properties, with a few features from the proximity with the F-layer.

\begin{figure*}[t]
\center{\includegraphics[width=0.99\linewidth]{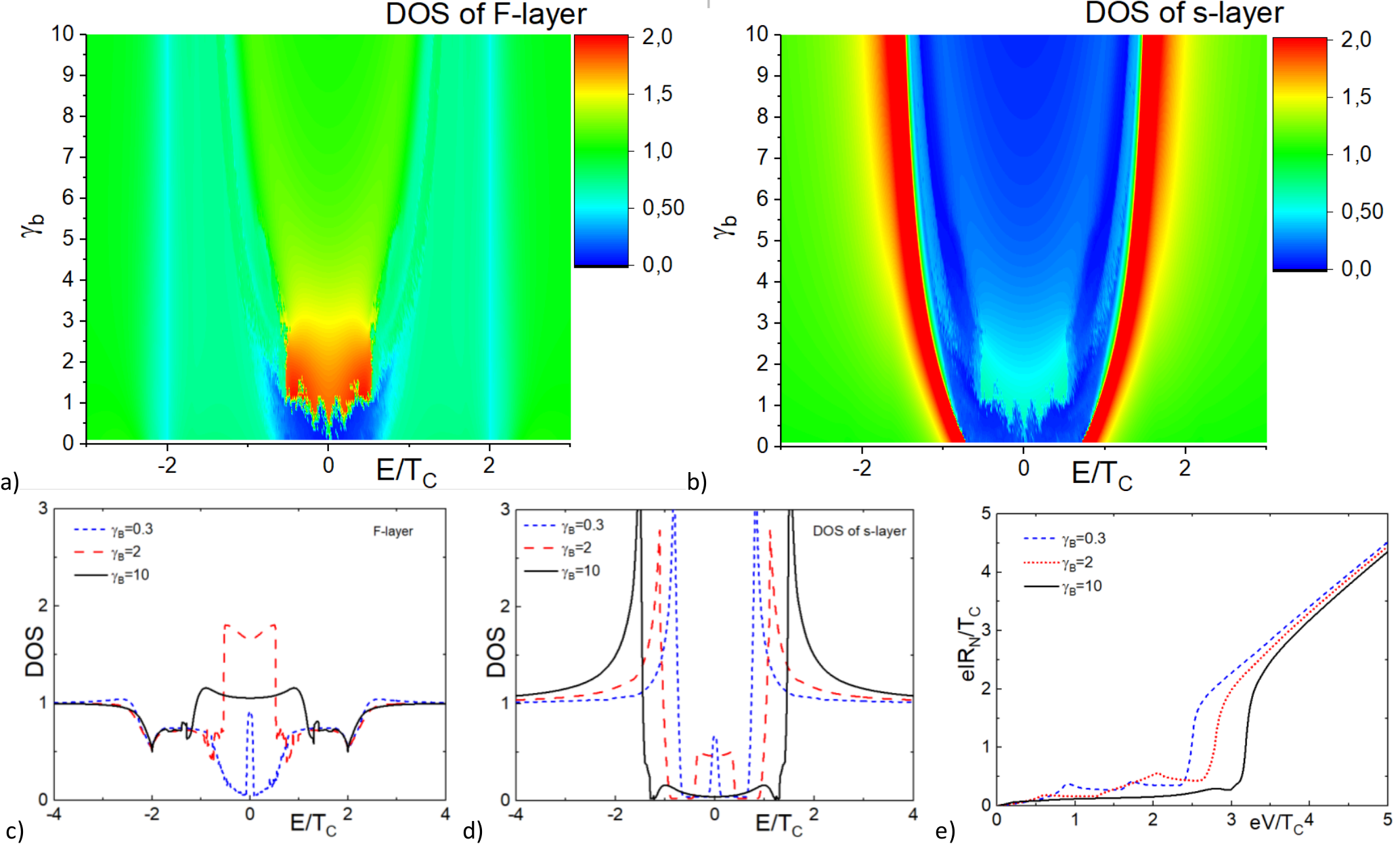}}
\caption{ DOS of sFS structure with intermediate thickness of  F- and s-layers ($d_F=1\xi$, $d_s=2\xi$)  and exchange energy $H=2 T_C$ a) in the middle of the F-layer and b) at the surface of the s-layer versus interface parameter $\gamma_B$; c,d) DOS in the middle of the F-layer c) and at the free surface of the s-layer d) for transparent boundary $\gamma_B=0.3$ (blue short-dashed  line), boundary with intermediate transparency $\gamma_B=2$  (red dashed line), and hard boundary  $\gamma_B=10$  (black solid line) exchange energies; e) IV-curves of the related SIsFS junction with small, intermediate and large thickness of the s-layer.   The other parameters are $\protect\gamma_B=0.3$ and $T=0.5 T_C$.   }
\label{gamma_B}
\end{figure*}

\subsection{Thin F- and s-layers, $d_F=0.5\xi,$  $d_s=0.6\xi$  \label{ThinF2}}

The DOS properties at the surface of the sFS structure are significantly different when both s- and F-layers are thin (See. Fig.\ref{H_sm_sm}). In this case, the proximity between both layers s and F, as well as penetrating proximity from the bulk S- layer are manifested in the most significant way. Actually, in the F-layer, DOS has all typical regions of parameters  (Fig.\ref{H_sm_sm}a,c): at $H<0.5 T_C$  it has the wide proximity induced gap, which is closing and reopening at $H>0.5 T_C$, at $H\approx3T_C$ the subgap zone appears and finally at $H=7 T_C$ it completely closes the gap. It is interesting, that in the interval of $H$ between $3T_C$ and $5T_C$, the sub-gap zone on DOS($H$) has a shape of the trident with 3 peaks. Such behavior of DOS is similar with studied triple-peak DOS in the F-layer of SFIFS structures \cite{Vasenko2}.

The DOS of the s-layer demonstrates nontrivial properties in the region of small $H$ (Fig.\ref{H_sm_sm}b,d). In the interval $H<T_C$, the coherence peaks are wider than in the other regions, and sub-gap states appears in the vicinity of point of the gap closing $H\approx0.3T_C$. Such properties appear due to effective proximity from the F-layer. At the higher exchange energies, the ''trident'' also appears in the s-layer, but as usual its amplitude is smaller than unity. The presence of the subgap states also leads to additional features on CVC of the junctions (Fig.\ref{H_sm_sm}e). As instance, at $H=5.2T_C$ there is additional peak on IV curve, which appears at $eV/T_C\approx 2.3$ and corresponds to the presence of subgap plateau which can be clearly seen on panel d) in the same figure. At the same time at the  $H=3T_C$ there is a peak at $eV/T_C\approx 0.7$, which corresponds to the difference between the bulk gap and minigap of the s-layer. This peak is absent in the cases of very small and very large exchange: in the former case the coherence peak is around bulk value due to strong effective proximity with the S-layer, and in the latter case the self-superconductivity is almost suppressed in the s-layer, and position of the peak is determined again by the bulk electrode.

%\begin{figure*}[t]
%\center{\includegraphics[width=0.9\linewidth]{L_F=0p5p3p1.pdf}}
%\caption{ DOS of sFS structure with thin thickness of F-layer ($d_F=0.5\xi$) and exchange energy $H=T_C$ a) inside %middle of F-layer and b) on the surface of the s-layer versus its thickness $d_s$; c,d) DOS in the middle of the F-layer c) and on %the free surface of the s-layer d) for small $d_s=0.5\xi$ (blue short-dashed  line), intermediate$d_s=2\xi$  (red dashed line), and %large $d_s=8\xi$  (black solid line) exchange energies; e) IV-curves of the related SIsFS junction with small, intermediate and %large thickness of s-layer.   The other parameters are $\protect\gamma_B=0.3$ and $T=0.5 T_C$.   }
%\label{ds_sm_1}
%\end{figure*}

%\begin{figure*}[t]
%\center{\includegraphics[width=0.9\linewidth]{L_F=0p5p3p3.pdf}}
%\caption{ DOS of sFS structure with intermediate thickness of F-layer ($d_F=0.5\xi$) and exchange energy $H=4T_C$ a) %inside middle of F-layer and b) on the surface of s-layer versus its thickness $d_s$;  The other parameters are $\protect%\gamma_B=0.3$ and $T=0.5 T_C$.   }
%\label{ds_sm_4}
%\end{figure*}

\subsection{The vicinity of the 0-$\pi$ transition}

%However, the above results don't demonstrate any properties of 0-$\pi$ transition. In the former case $d_F=2\xi$, there is no support of bulk superconductor in s-layer and DOS becomes normal-metal-like at the interesting region of parameters relating to 0-$\pi$ transition. In the latter case $d_F=0.5\xi$  0-$\pi$ transition requires too large value of exchange energy, which also leads to disappearing of the effects.
 To reveal the properties of SIsFS structure at the point of  0-$\pi$ transition we consider the case with thickness of ferromagnetic layer $(d_F=\xi).$ At this thickness  significance of penetrative proximity from the S bulk electrode to the s-layer does not too strong  and  allows to realize  the 0-$\pi$ transition at $H=11T_C$. In Fig.\ref{0pitr}d we plot the spatial distribution of the real part of anomalous Green function as a function of exchange energy $H$. It is seen that  increase of the exchange energy really provides the change of the sign of $Re F$ in the thin s-layer in the vicinity of  $H=11T_C$. This transition has small influence on density of states inside ferromagnetic layer (Fig.\ref{0pitr})a, because it has almost normal-like shape.  
%as well as in the other structures with high exchange field. 
At the same time, the $0-\pi$ transition can be detected from the DOS on the surface of the thin s-layer, as it shown in Fig (\ref{0pitr})b,d,e. As in the previous cases, the DOS of that system has entire gap at the small $H$, the finite sub-gap zone  in the interval of exchange fields $(2T_C, 5T_C)$, and gapless states at the higher exchange energies. Note that even in the gapless state there is a minor decrease of the DOS around zero energy.  The value of this feature is sensitive to the 0-$\pi$ transition, as it shown on the enlarged picture of DOS in Fig.\ref{0pitr}d. At the point of the 0-$\pi$ transition, that decrease disappears, and restores again in the region of the $\pi$-state with almost the same deepness of the zero energy feature (Fig. \ref{0pitr}e). This change provides slightly more straight IV curve of SIsFS junction at the point of 0-$\pi$ transition (Fig (\ref{0pitr}f), but this change is insufficient for detection of the transition by consideration of the resistive branch of the CVC. 
%At the same time, the measurements of the local surface DOS of the s-layer probably can reveal the 0-$\pi$ transition.

\subsection{Influence of the SF interfaces}

%The  variations of the interface parameter at the interface between superconductor and ferromagnetic metals can be also the reason of the inhomogeneity of the DOS in hybrid structures. 
We consider the influence of the interface parameter $\gamma_B$ on DOS of SFs structure with $d_F=\xi$ and $d_s=2\xi$ (See Fig. \ref{gamma_B}). The shown dependencies reveal the competition between the three processes in the system. The first one is the penetrating proximity from the bulk S-layer. In the case of extremely transparent interfaces ($\gamma_B \rightarrow 0$), there is domination of proximity induced correlations, which support superconductivity in the bi-layer and provide the well-formed gap in the both ferromagnetic and thin superconductive s-layer.  However, the appearance of the finite boundary resistance weakens the impact of that process. Then, in the interval of $\gamma_B$ between 1 and 3, the proximity between s- and F- layers has a main role, leading to formation of the significant sub-gap zone. Finally, at the very large resistance of interface ($\gamma_B > 10$), the s-layer is protected from the proximity from the S- and F- layer, and the DOS in it gradually restores the properties of a bulk superconductor. However, the sub-gap states still noticeable even at $\gamma_B=10$ (Fig. \ref{gamma_B}d). 

In terms of CVC such changes of $\gamma_B$ are mostly defined by the change of the position of the coherence peak, leading to the shift of the current drop (Fig. \ref{gamma_B}e). However, small additional peaks provide some information about the presence of the subgap zone and  about an actual value of the pair potential in the s-layer.

It also should be noted, that in the case of very thin s- and F- layers, the proximity effect can provide significant increase of the gap in the case of the small  $\gamma_B$ value. It can be realized, when the induced minigap in the bi-layer is larger than the self-gap of s-layer in such system.

\section{Discussion}
In this paper we have studied the influence of the proximity effect in the SIsFS junctions on the density of states in the vicinity of the tunnel barrier and on the resistive branch of current-voltage characteristics. We have demonstrated that the DOS shape at the Is interface  significantly depends on the thickness of the s- and F- layers, exchange energy of ferromagnetic metal, and interface parameters.

Generally, all revealed features correspond to the interplay of three phenomena: penetration of superconducting correlations from the bulk S-layer into the Fs-bilayer, mutual interaction between the s- and F- layers and formation of intrinsic superconductivity in the thin s-layer.

%The most interesting features appear in the case of small exchange exchange energy in order of superconducting gap. 
The most interesting effects occur in the case of small values of the exchange energy, comparable to the superconducting pair potential.
This regime provides the formation of the subgap zones, which lead to the appearance of additional peaks on IV-curves. 

Actually, the measurement of CVC in SIsFS junction may serve as an effective source of information about the electronic structure inside the s-layer. There are two peaks on IV-curves: the first one appears at the smaller voltage and corresponds to the difference between the position of coherence peak inside s-layer and the gap of bulk superconductor. This peak has weak dependence on the parameters like $H$ or $\gamma_B$, and it should be non sensitive to sample parameters variations. At the same time, the value of $\Delta$ can be also obtained from the value of critical current of the junction, which provides possibility to cross-check the estimation of main peak position in the single experiment.

The second peak on IV curve corresponds to the presence of subgap states, induced into the s-layer from a ferromagnetic metal.  These subgap states are sensitive to variation of junction parameters, so the that peak position can be shifted in the different samples.  This peak usually appears near the current drop on IV-curves. 

We have also found the influence of the 0-$\pi$ transition on the DOS of s-layer. While in the 0 or $\pi$ states there is a significant deep on the DOS, in the vicinity of the transition it becomes almost normal-metal like. Even though it doesn't provide any significant effect on CVC of SIsFS Josephson junction, this dip can be measured in sFS structures with free surface of thin s-layer by scanning tunneling spectroscopy technique \cite{Stolyarov1, Stolyarov2}. 

\textbf{Acknowledgments.} The authors acknowledge fruitful discussions with
V. S. Stolyarov, V. V. Bol'ginov and D. Massarotti. The general study of the DOS and IV-curves was
supported by RFBR (18-32-00672 mol-a), numerical calculations of the systems with thin layers were done
with support of Russian Science Foundation (20-69-47013) and study of zero-energy peaks was supported by RFBR (19-52-50026). A.N. also acknowledges scholarship of the BASIS foundation.

\end{document}